# ABORDAREA IERARHIZATA A DECIZIILOR DE GRUP

**MARIAN DRAGOI,**

**CIPRIAN PALAGHIANU**

Facultatea de Silvicultură a Universităţii „Ştefan cel Mare" din Suceava

**Group decision makers making process - an analytic hierarchy approach**

Abstract:

The paper deals with a step-wise analytic hierarchy process (AHP) applied by a group of decision makers wherein nobody has a dominant position and it is unlikely to come to terms with respect to either the weights of different objectives or expected utilities of different alternatives. One of the AHP outcomes, that is the consistency index is computed for each decision maker, for all other decision makers but that one, and for the whole group. Doing so, the group is able to assess to which extent each decision maker alters the group consistency index and a better consistency index could be achieved if the assessment procedure is being resumed by the most influential decision maker in terms of consistency.

The main contribution of the new approach is the algorithm presented in as a flow chart where the condition to stop the process might be either a threshold value for the consistency index, or a given number of iterations for the group or decision maker, depending on the degree to which the targeted goal has been decomposed into conflictual objectives.

**Introducere**

Gestionarea durabilă a resurselor naturale presupune, de cele mai multe ori, aşa-numitul management participativ care, la rândul lui, se rezumă la deciziile multicriteriale de grup. Metoda proceselor analitice ierarhizate (PAI)[3] a fost creată de profesorul Thomas Saaty [6] și, în prezent, se bucură de un interes crescând, datorat și faptului că a fost implementată într-un produs informatic foarte flexibil, respectiv Expert Choice.

În silvicultură, metoda a făcut obiectul a două aplicații ale proceselor analitice ierarhizate [2,3], urmate apoi de un interesant studiu de caz asupra modului în care preocupările de conservare a biodiversității se regăsesc în fundamentarea deciziilor privind amenajarea pădurilor [4]. O altă aplicație interesantă a fost identificarea punctelor tari, a celor slabe, a oportunităților și pericolelor (analiza SWOT[4]) asociate certificării pădurilor [5].

Într-o aplicație mai complexă [1], pe primul nivel se situează *gestionarea durabilă* a pădurilor, pe nivelul doi grupurile de interese de care depinde realizarea respectivului obiectiv, pe nivelul al treilea mai multe criterii de analiză iar pe nivelul al patrulea politicile forestiere alternative.

Metoda constă de fapt în *ierarhizarea procesului de decizie*, urmată de evaluarea tuturor componentelor de pe un anumit nivel în raport cu fiecare componentă de pe nivelul imediat superior (figura 1). Prin „componentă" se poate înțelege *scop final* dacă e vorba de primul nivel, „obiectiv" dacă ne referim la al doilea nivel, respectiv mijloc pentru nivelul trei.

---

[3] Termenul consacrat in literatură este Analytic Hierarchy Process (AHP).
[4] SWOT este un acronim pentru „puncte tari" (*strenghts*, în eng.), puncte slabe"(*weaknesses*), oportunități (*opportunities*) și pericole (*threats*), considerate a fi caracteristici ale oricărei decizii pe termen lung.

## Scopul cercetărilor

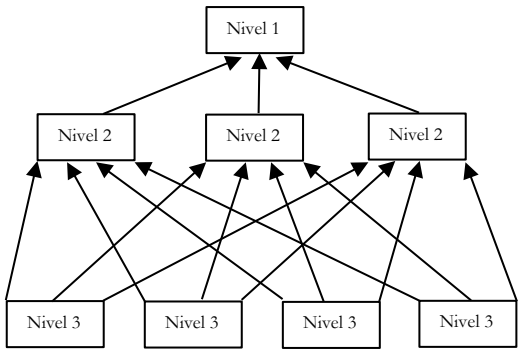

**Figura 1 Principiul analizei ierarhizate a proceselor**

Deşi metoda PAJ permite şi determinarea gradului de coerenţă a evaluărilor ce fundamentează o anumită decizie, atunci când sunt mai mulţi decidenţi, cu poziţii (ponderi) egale, este foarte posibil ca indicele de coerenţă la nivel de grup să fie nesatisfăcător, datorită faptului că fiecare membru al grupului evaluează diferit ponderile criteriilor şi/sau efectele alternativelor decizionale, în raport cu fiecare criteriu în parte.

Motivul unei evaluări diferite nu este neapărat un interes anume, ci poate fi, la fel de bine, o percepţie diferită a riscurilor asociate procesului decizional în sine: de pildă, distrugerea habitatelor umede este percepută într-un anume fel de un biolog, şi altfel de cel ce construieşte un drum forestier: primul va avea, prin natura profesiei o atitudine prudentă în raport cu orice zonă umedă, al doilea, din contra, una de acceptare a riscului (figura 2).

În astfel de situaţii aparatul matematic pe care se bazează metoda - şi, în general, orice altă tehnică de fundamentare a deciziilor - devine inutil. Problema ce se pune într-un astfel de context este aceea a *armonizării evaluărilor individuale*, astfel încât coerenţa evaluării colective să fie ameliorată; altfel spus, procesul analitic ierarhizat să devină un mijloc de *negociere* între factorii de decizie implicaţi.

O evaluare coerentă (fie a ponderii criteriilor de decizie, fie a utilităţii fiecărei variante decizionale în raport cu fiecare criteriu în parte) se realizează atunci când orice combinație de trei criterii/variante respectă următoarea condiție logică:

Dacă $w_i > w_j$ și $w_j > w_k$, atunci $w_i > w_k$ pentru $w_{i,j,k}$; $i,j,k = 1, \ldots, n$ „n" fiind numărul de criterii sau variante decizionale iar w ponderea criteriilor sau utilitățile variantelor în raport cu un criteriu anume. Dar dacă w reprezintă valori medii pe decidenți, este posibil ca respectiva condiție să nu mai fie îndeplinită, ceea ce face necesară negocierea între decidenți.

**Rezultate**

Nefiind vorba de o cercetare experimentală propriu-zisă, nu mai este oportună discutarea metodei de investigație (firește, simularea), și cu atât mai puțin a materialului faptic care să dovedească validitatea sau invaliditatea ipotezelor de la care s-a pornit. În figura 3 este prezentată schema logică a ceea a fost denumită analiza ierarhizată în trepte, ce se bazează pe un proces iterativ de re-evaluare a criteriilor, respectiv a variantelor decizionale, astfel încât să se asigure cel puţin creşterea gradului de coerenţă a deciziei finale.

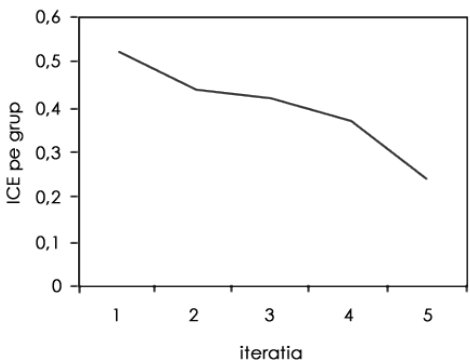

**Figura 3 Descreșterea indicelui de consecvență al grupului prin analiza ierarhizată în trepte**

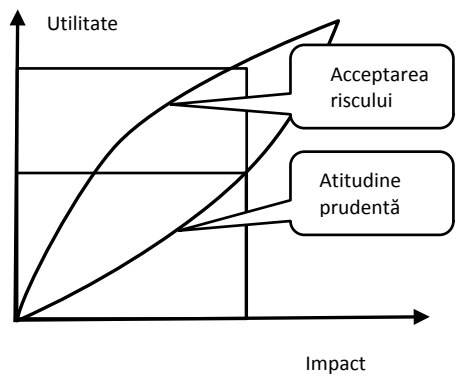

**Figura 2 Percepții diferite asupra riscului conduc la evaluări**

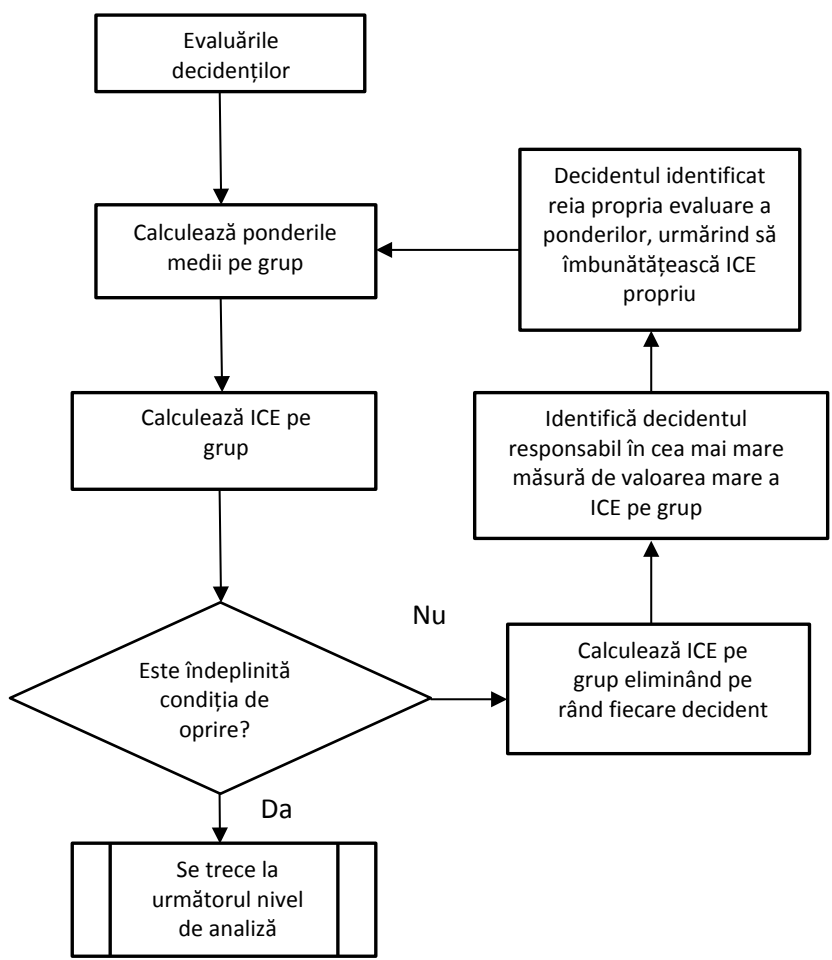

**Figura 4  Schema logică a utilizării analizei ierarhizate în trepte**

Prezentarea detaliată a metodei analizei ierarhizate poate fi găsită în toate lucrările citate în bibliografie, dar ceea ce trebuie subliniat în acest context este faptul că metoda se bazează exclusiv pe cinci trepte de evaluare calitativă ce corespund lingvistic unor formulări de tipul „*extrem de important (comparativ cu .. .)*", „*foarte important*", „*mult mai important*", „*mai important*" şi „*la fel de important*". Valorile numerice corespunzătoare sunt 9,7,5,3,1, fiind admise şi note intermediare (8, 6, 4, 2).

Indicele de coerenţă a evaluării trebuie să fie mai mic de 0,1 şi se calculează raportând un indice specific fiecărei situaţii decizionale la un alt indice, tabelat, ce corespunde unor alegeri complet aleatoare, deci unor decizii complet incoerente.

Algoritmul prezentat în figura 3 a fost implementat într-un program de calculator, ce a fost apoi utilizat la simularea mai multor decizii de grup, în situaţii potenţial conflictuale. Rezultatul unei astfel de simulări, publicate anterior [2] este prezentat în figura 4. Se observă că după cinci iteraţii indicele de consecvenţă a scăzut de la 0,5 la 0,25. Faptul că nu s-a ajuns la o valoare optimă (cel mult egală cu 0,1) nu diminuează cu nimic importanta unei astfel de abordări întrucât orice decizie de grup scoate în evidenţă antagonismul potenţial dintre obiective, pe de o parte, şi dintre consecinţele variantelor decizionale, pe de altă parte.

**Concluzii si discuţii**

În condiţiile unor decizii generate de situaţii conflictuale sau care pot genera situaţii conflictuale, managementul participativ riscă să rămână lipsit de conţinut, în condiţiile în care nu există metode adecvate de fundamentare a deciziilor. *Cu cât creşte numărul criteriilor creşte şi posibilitatea evaluării incoerente a importanţei acestora.* O soluţie ar fi aceea a descompunerii criteriilor în subcriterii, dar aceasta conduce la creşterea gradului de complexitate a problemei, fapt ce face şi mai necesară utilizarea unei metode fundamentate matematic. Simulările şi testele făcute pe mai multe persoane au dovedit că evaluări coerente logic se obţin de prima dată ori de câte ori numărul obiectivelor sau variantelor este trei. Având în vedere faptul că variantele urmează a se compara două câte două în raport cu fiecare obiectiv în parte, se poate aprecia că deşi pot fi urmărite mai mult de trei obiective este bine ca numărul variantelor de analizat să se limiteze la trei, *deoarece este mai importat să se asigure un ICE mai mic de 0,1 la evaluarea finală decât în faza preliminară, de ponderare adecvată a obiectivelor.*

Metoda, în forma iniţială, nu înlătură complet riscul unor discuţii interminabile între membrii unui colectiv de decizie; în forma îmbunătăţită (figura 3) aduce intr-o astfel de discuţie o bine-venită nota de pragmatism, dându-i fiecărui membru posibilitatea să-şi evalueze continuu coerenţa propriilor evaluări şi efectul pe care propriile-i opinii le are asupra deciziei finale.

Un alt aspect, evidenţiat în faza de testare a programului informatic este echilibrul care ar trebui să existe de la bun început între numărul obiectivelor convenite şi numărul experţilor ce fac parte din grupul de decizie, dublat de disponibilitatea respectivilor factori de decizie de a reveni asupra propriilor evaluări.

**Bibliografie**